# Layer-by-Layer assembling and characterizations of dye-polyions onto solid substrate by Electrostatic adsorption process

Md. N. Islam, P.K. Paul, S.A. Hussain and D.Bhattacharjee*
Department of Physics, Tripura University, Suryamaninagar-799130, Tripura West, INDIA

**Abstract:**
Self assembled films of organic azobenzene dye chicago sky blue 6B (CSB) have been fabricated onto solid substarte by electrostatic alternate adsorption of polycation ploy(allyl amine hydrochloride) (PAH) and CSB. UV-Vis absorption spectroscopic studies reveal the successful incorporation of CSB molecules into LbL films and consequent formation of aggregates. This view is supported by FTIR spectroscopic studies. Scanning electron microscope picture confirms the formation of nano crystalline aggregates in the LbL films. About 15 minute is required to complete the electrostatic interaction between PAH and CSB molecules in one bi-layer LbL film.

## Introduction:

Multilayer films of suitable organic compounds have been studied for more than 60 years because they allow fabrication of multicomposite supramolecular assemblies of tailored structure [1,2]. The well structured molecular assemblies of organic functional molecules on a nanometer scale may be achieved by Langmuir-Blodgett and Layer-by-Layer Self Assembled monolayer techniques [3, 4]. However the most recent method along this direction is thin film deposition based on Electrostatic Layer-by-Layer adsorption of oppositely charged species on to solid substrates. This unique technique covers a wide diversity of materials and film fabrication is performed under mild condition which is extremely important for preserving activity of biomolecules [5]. Electrostatic Layer-by-Layer method has proven to be quite effective for the layering of biological molecules such as proteins, enzymes, DNA, Cell membrane and viruses [6, 7]. Recently Kunitake and co-workers have shown that low molecular weight chromophoric dyes are also capable of spontaneous assembly in aqueous solutions where suitable charged surfaces are formed on the substrates [8]. Cooper and others also demonstrated that anionic azo dyes could be assembled when interacting with a cationic polymer on to a solid support [9]. More significantly dye-containing Layer-by-Layer assemblies may be utilized for various potential optical and sensor applications [1, 10]. It is therefore extremely important to examine in detail the behaviour of charged small molecules dyes as the basic components for the Layer-by-Layer self assembled films. In particular it is also of fundamental importance to correlate the dimension, charge density, solubility, nature of aggregation etc. for these molecular assemblies on a nanometer scale.

Small molecule azobenzene dyes are now receiving much attention among the scientists and researcher as they are being used in colorants, stains and markers [11]. These dyes generally contain two or more azo groups which have sulfonated groups that provide solubility in water. They are now also being used in a developing range of biochemical assays for amyloid and other proteins and they also have potential for pharmaceutical use in treatment of HIV and neurodegenerative diseases such as Alzheimer's and Scrapie [12, 13]. This class of dyes is mostly to form dimmers or higher order aggregates in solutions [14] as a result of dye-dye intermolecular interactions. So in this regard it also crucial to understand their assembling mechanism, aggregation behaviour when deposited onto solid substrates by using Electrostatic Layer-by-Layer self assembly method. It is relevant to mention here that LbL method with the possibility of control at the molecular level is used to assemble azobenzene dyes with functional polymer films and is expected to gain significant importance [15]

There are few reports on the formation of LbL self assembled films of organic dye molecules [9, 17]. Either the molecules had alkyl chains [16] or a correct combination of polyion had been used [17]. Due to the presence of less number of charged groups in these molecules physiadsorption of such molecules on to solid substrates is not straight forward. Material loss occurs and is substantive by washing the film with water in many cases.

In this present article we have examined the alternate multilayer adsorption of charged azobenzene dye molecules Chicago Sky Blue 6B (anionic) with poly(allylamine hydrochloride) (polycationic) by Electrostatic Layer-by-Layer technique. The molecular assembly process of the polycation-dye pairs was investigated by FTIR, UV-Vis absorption and steady state Fluorescence spectroscopy. Scanning Electron Microscopy (SEM) provides the information about the domain structures of the dye molecules in the LbL films.

**Experimental:**

The anionic sulfonated azobenzene dye used in these experiments is Chicago Sky Blue 6B (CSB) or Direct Blue 1. Poly(allylamine hydrochloride)(PAH) is used as polycation for Layer-by-Layer deposition of the dye molecules onto quartz substrates. CSB (Molecular weight 992.82, purity >99%) and PAH (Molecular weight 70,000, purity>99%) were purchased from Aldrich Chemical Co., USA and handled using proper safety procedures [18]. They were used without further purification and chemical structures are shown in figurer 1a. Aqueous solutions of different concentrations of CSB (ranging from $10^{-4}$M -$10^{-8}$M) were prepared using the triple distilled deionised Milli-Q water with 18.2 MΩ resistivity.

Layer-by-Layer Self Assembled films have been prepared by dipping rigorously the cleaned fluorescence grade quartz substrates alternately into polymer solution and dye solutions utilizing the electrostatic interactions between the oppositely charged species and the Vanderwalls interaction between the substrate and polymer[19]. Initially the quartz substrates were immersed in PAH solution for 15 minutes followed by rinsing in water bath for two minutes. This rinsing actually washes off the surplus cation attached to the substrates. The substrate was then immersed on the dye solution for 15 minutes followed by same rinsing procedure in a separate water bath. As a result one bilayered of CSB-PAH film was formed onto the quartz substrate. The multilayered LbL films were prepared by repeating the whole sequence up to a desired number. pHs of dye solutions were adjusted with NaOH and $CH_3COOH$.

For photophysical characterizations of the CSB-PAH LbL films we used UV-Vis Absorption Spectrophotometer (Lambda25, Perkin Elmer) Fluorscence Spectrophotometer (LS55, Perkin Elmer). The FTIR spectra were recorded using a Bruker IFS 66/v FTIR spectrometer. The Scanning Electron Micrograph of the LbL films were obtained by using High resolution Field Emission Scanning Electron Microscope installed at Katholeike Universitit Leuven, Belgium.

**Results and discussions:**

Figure 1 shows the UV-Vis absorption spectra of the aqueous solution of Chicago Sky blue 6B (CSB) of $10^{-5}$M concentration of CSB along with CSB microcrystal and 1 bi-layer CSB-PAH LbL film.

From the figure it is observed that in all the cases, intense broad longer wavelength band is observed in the 450-750 nm region along with a prominent high energy band in the 275-350 nm region and is consistent with the reported literature [4]. Moreover, in solution absorption spectrum a weak hump is observed with peak at around 400 nm. The longer wavelength broad band has intense peak at 618 nm in solution. Whereas, in CSB microcrystal this peak is blue shifted to 597 nm. In LbL film this peak coincides with the solution absorption spectrum. Moreover the absence of high energy weak hump at 400 nm in the microcrystal as well as in LbL film spectra indicates reabsorption effect owing to the formation of aggregates. Moreover, no shift of longer wavelength band in comparison to solution spectrum, in case of LbL film and a blue shift of this band in case of microcrystal clearly indicates that the nature of aggregation is different in microcrystal and in LbL film. This is happened due to change in orientation of molecules in the aggregates.

Figure 2 shows the FTIR spectra of CSB-PAH LbL film along with CSB in KBr pallet for comparison. In both the spectra a band at 1610 $cm^{-1}$ is observed which can be attributed to the bending vibration of $–NH_2$ group of CSB [20]. Moreover, in CSB-PAH LbL film spectrum, the bands due to the stretching vibration of O-H (COOH) and N-H overlap and shifted to a higher wave number 3515 $cm^{-1}$ compared to that of CSB in KBr pallet spectrum at 3413 $cm^{-1}$. This shows that CSB interacts electrostatically with PAH in the LbL films.

High resolution scanning electron micrograph (SEM) of 10 bi-layer CSB-PAH LbL film is shown in figure 3. This picture clearly shows the nanodimensional aggregates of dye-polymer complex with sharp and distinct edges. Thus the SEM photograph provides the compelling visual evidence of aggregation in LbL films.

UV-Vis spectroscopy was primarily used to investigate the assembly process of dye layers and aggregation phenomena [21, 15]. Figure 4a. Shows UV-Vis absorption spectra of Layer-by-Layer self assembled films of CSB-PAH onto quartz substrate as a functions of layer number (1 to 20 alternate layer of PAH and CSB). The dye/polycation deposition time for all the cases were taken as 15 minutes. The absorption spectra of different layered LbL films show similar band profile irrespective of Layer number except an increase in intensity distribution of the peaks observed at 321 nm and 618 nm in UV and visible region of the spectra respectively. From the inset of figure 4a it is observed that the intensitie of absorbance at $\lambda_{max}$ increase linearly in proportion to the number of dye-polymer pair layers. This signifies that charge compensation (reversal) during adsorption was sufficient to proceed with each subsequent deposition. This also reveals that the CSB-PAH complex species increases with the increase in film thickness and definitely confirms the successful incorporation of CSB molecules in the PAH-CSB LbL films. The films were clear transparent and uniform with good optical quality. The $\lambda_{max}$ values attributed mostly to the $\pi$-$\pi^*$ transition for the dye-poycation pairs [4]. The change in $\lambda_{max}$ for PAH-CSB LbL films with respect to CSB aqueous solution remained same with increasing layer number. This certainly indicates that the degree of aggregation also remained unaltered with increasing layer thickness.

Fig 4b represents the normalized fluorescence emission spectra of different layered (1 to 10 bilayers) CSB-PAH LbL films along with CSB microcrystal spectrum for comparison. The excitation wavelength is 600 nm. The spectra show the distinct and prominent band systems with peaks at 693 nm and 720 nm. However the most interesting thing in this observation is that the broad and structureless band in the 825-875 nm region of the emission spectra changes its intensity distribution among various bands within itself. This broad band in the higher wavelength side is merely independent of the number of layers and which is quite similar to that of the CSB microcrystal emission spectrum. The origin of this band is mainly due to the closer association of the CSB molecules and the subsequent deformation produced in their electronic levels. This behavior is also observed in some low molecular weight dyes [26]. Moreover the 720 nm band in cases CSB microcrystal spectrum is significantly broadened due to the closer association of the CSB molecules and overlapping of the several vibrational states in microcrystal.

Figure 5 show the UV-Vis absorption spectra of 1 bi-layer CSB-PAH LbL film for different dye deposition times ranging from 1 min to 60 min. From the figure it is observed that the intensity of the absorbance of the films increases with increasing dye deposition time. The absorbance intensity is maximum when the dye deposition time is 15 min and intensity remains constant on further increasing the dye deposition time. This is also manifested in the plot of the intensity of the absorption maxima ($\lambda_{max}$) versus deposition time as shown in the inset of figure 5. This reveals that the interaction of dye molecules with the positively charged PAH layer on quartz substrate was completed up within 15 minutes and after 15 minutes no PAH site remains free in the LbL film for further interaction with CSB molecules.

The effect of pH of the CSB solution in CSB-PAH LbL films has also been studied. Figure 5 shows the absorption spectra of 1 bi-layer CSB-PAH LbL films deposited using CSB solution of varying pH. The concentration of the CSB solution and deposition time was kept fixed at $10^{-4}$M and 15 minute for all the cases. From the figure it is observed that the main CSB absorption band show a small blue shift at higher pH although the spectral profile remains almost unaltered. This blue shifting of the main visible band may be due to the fact that the two acidic groups of the dye molecules are too remote to influence one another and certainly both half of the molecules are deprotonated at such high pHs [27].

Interestingly it is observed that the intensity of absorbance varies significantly with pH (inset of figure 6). The intensity of absorption bands is maximum when the pH of the solution is 6.8. However, the absorbance intensity decreases for low as well as high pHs. This may be due to the change in ionic nature of the CSB solution with change in pHs. The plot of the ratio of intensities of 618 and 321 nm band as a function of pH (inset of figure 6) show that these values are almost independent of pH. This indicates that the band at 618 nm cannot be due to any kind of intermolecular charge transfer and the nature of aggregation also remains unchanged with pH.

**Conclusion:**

In summary our result show that multilayer films of azobenzene dye CSB onto solid substrate (quartz) can be obtained by electrostatic layer-by-layer (LbL) self assembled technique. UV-Vis absorption spectroscopic studies reveal the successful incorporation and consequent molecular aggregation into LbL films. SEM photograph confirms the existence of nanodimensional aggregates into LbL films. Electrostatic interaction between the dye CSB and polycation PAH, resulting the formation of CSB-PAH complex into LbL films is demonstrated by FTIR spectroscopic studies. It is observed that almost 15 minute is required to complete this electrostatic interaction between dye and polycation.

**Acknowledgements:**

The authors are grateful to CSIR, Government. of India for providing financial assistance through CSIR project Ref. No. 03(1080)/06/EMR-II and.The authors are also thankful to the Centre for Surface Chemistry and Catalysis, Katholieke Universiteit, Leuven, Belgium for providing scanning electron microscope facility.

**Figure Captions:**

1a. Molecular structure of Chicago Sky Blue 6B.

1b. UV-Vis absorption spectra of aqueous solution CSB of different concentration along with CSB Microcrystal absorption spectra

1c. UV-Vis absorption spectra of 1 bilayered PAH-CSB LbL films deposited from CSB aqueous solution having different concentration of CSB.

2a. UV-Vis absorption spectra of different bilayered (1 to 20 alternate layer) PAH-CSB LbL films along with the variation of absorbance intensity of main visible peak with bilayer number (Inset), concentration of dye in solution = $10^{-4}$M

2b. Steady state fluorescence emission spectra of different bilayered PAH-CSB LbL films.

3a. UV-Vis absorption spectra of 1 bilayered PAH-CSB LbL films for different dye deposition time along with the variation of main visible peak with dye deposition time (Inset), Concentration of CSB in solution =$10^{-4}$M.

3b. Steady state fluorescence spectra of 1 bilayered PAH-CSB LbL films for different dye deposition time, Concentration of CSB in solution =$10^{-4}$M.

4a. UV-Vis absorption spectra aqueous solution of CSB for different pHs with Concentration of dye$10^{-5}$M

4b. UV-Vis absorption spectra of 1 bilayered PAH-CSB LbL films deposited from solutions of CSB having different pH along with the variation of UV and Visible peak with the pH of the dye solutions.

5. Scanning Electron Micrograph of 1 bilayered PAH-CSB LbL films deposited from $10^{-4}$M dye solution.

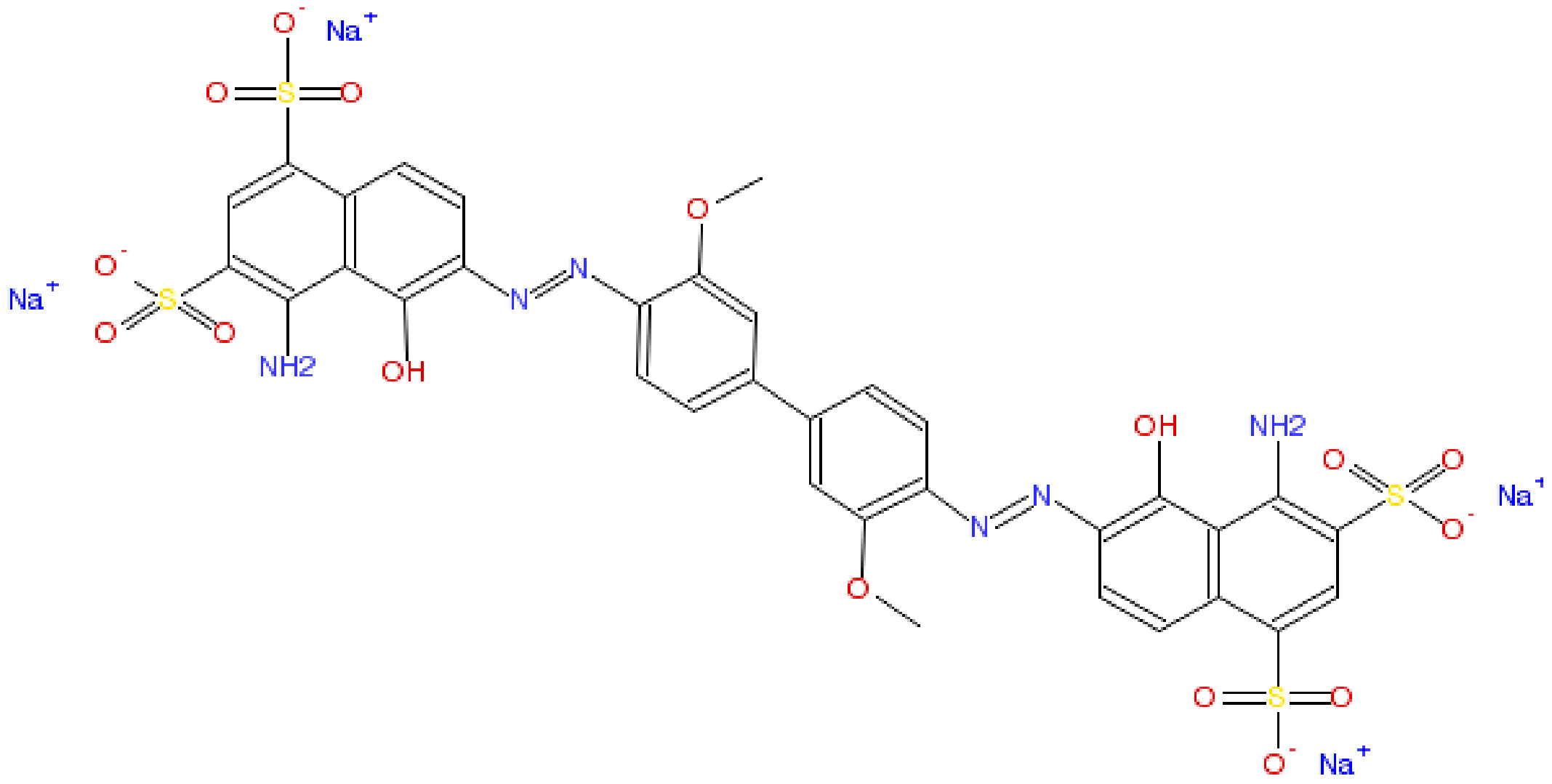

Figure 1a: Md. N. Islam et. al.

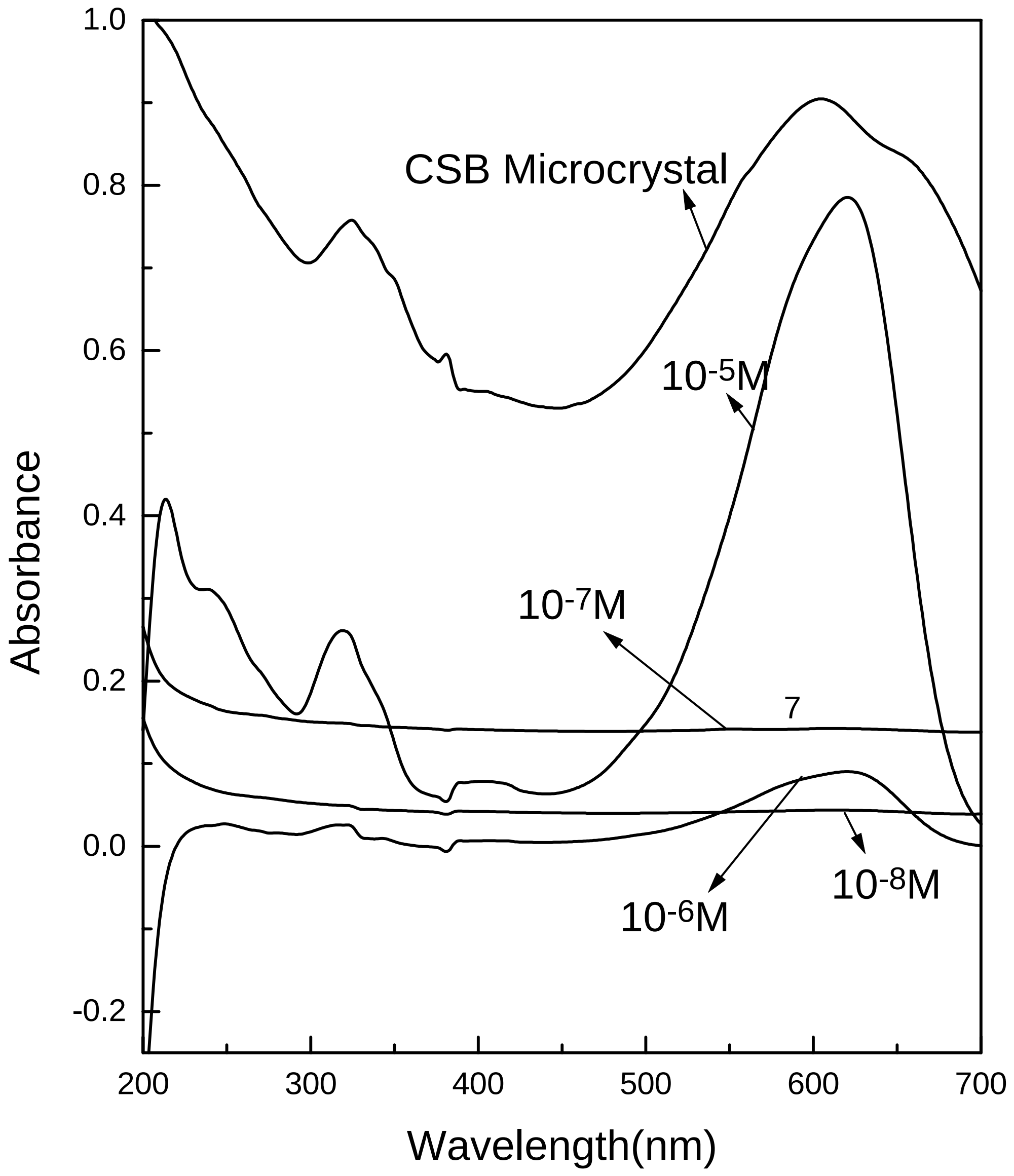

Figure 1b: Md. N. Islam et. al.

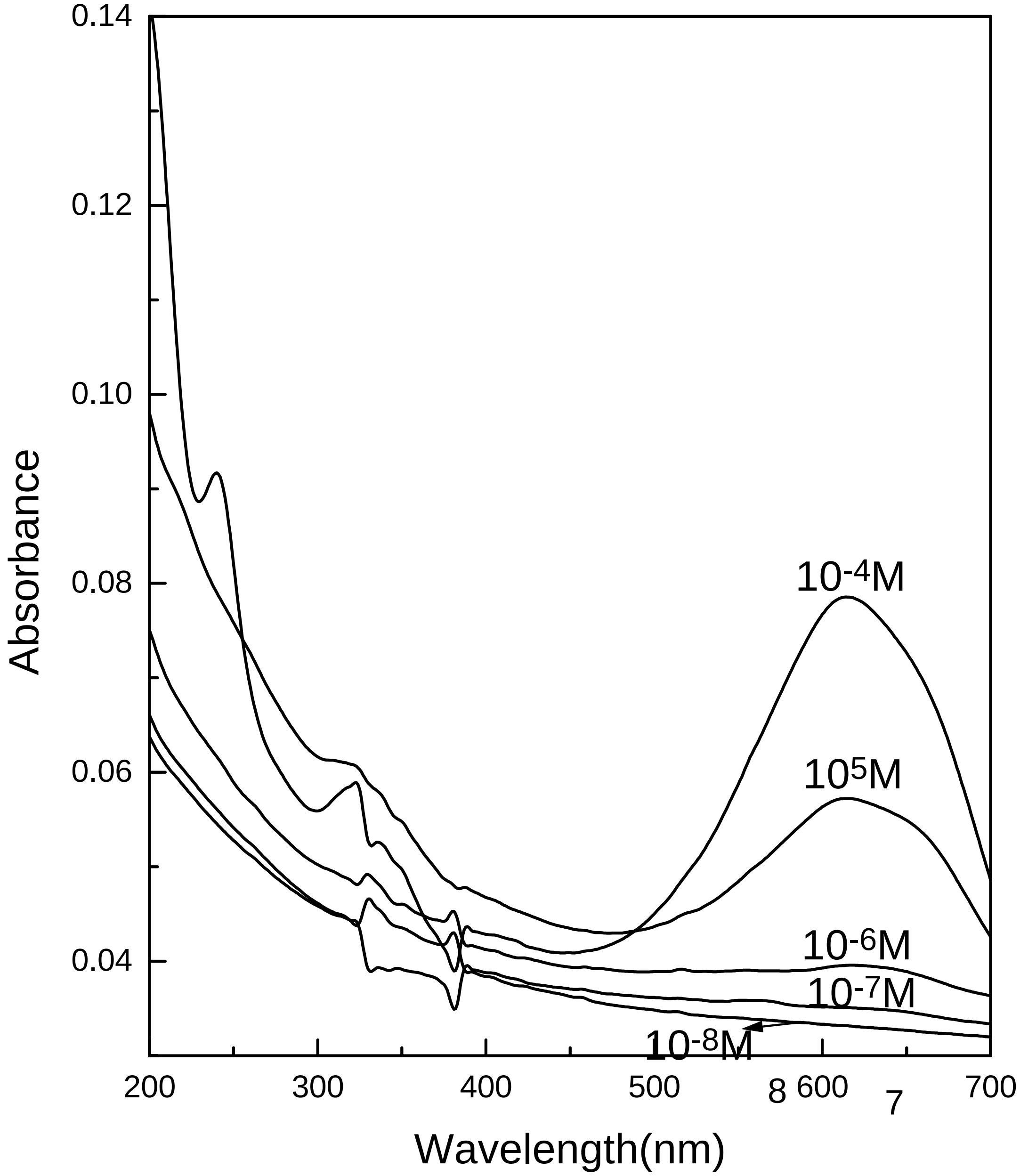

Figure 1c : Md. N. Islam et. al.

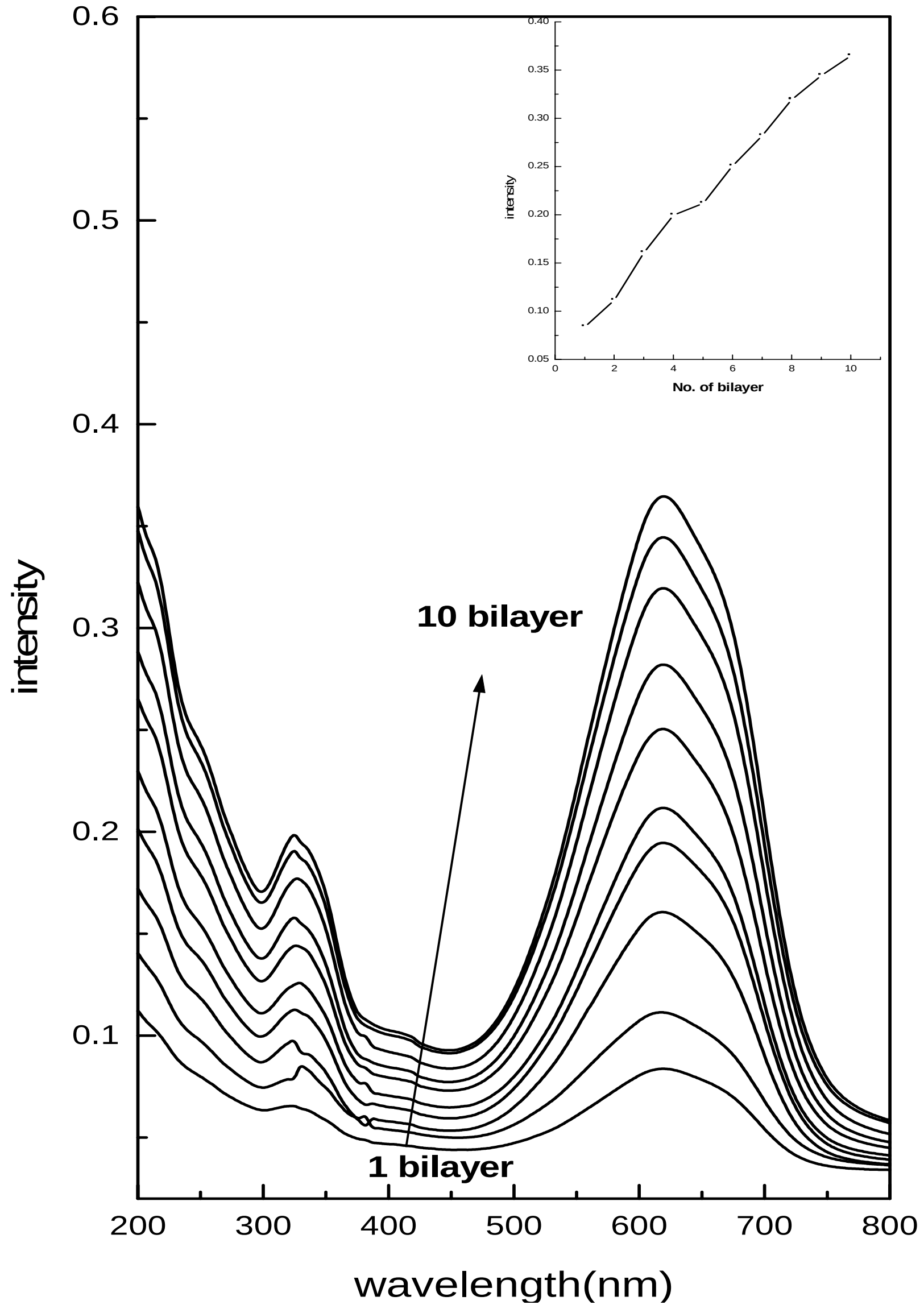

Figure 2a : Md. N. Islam et. al.

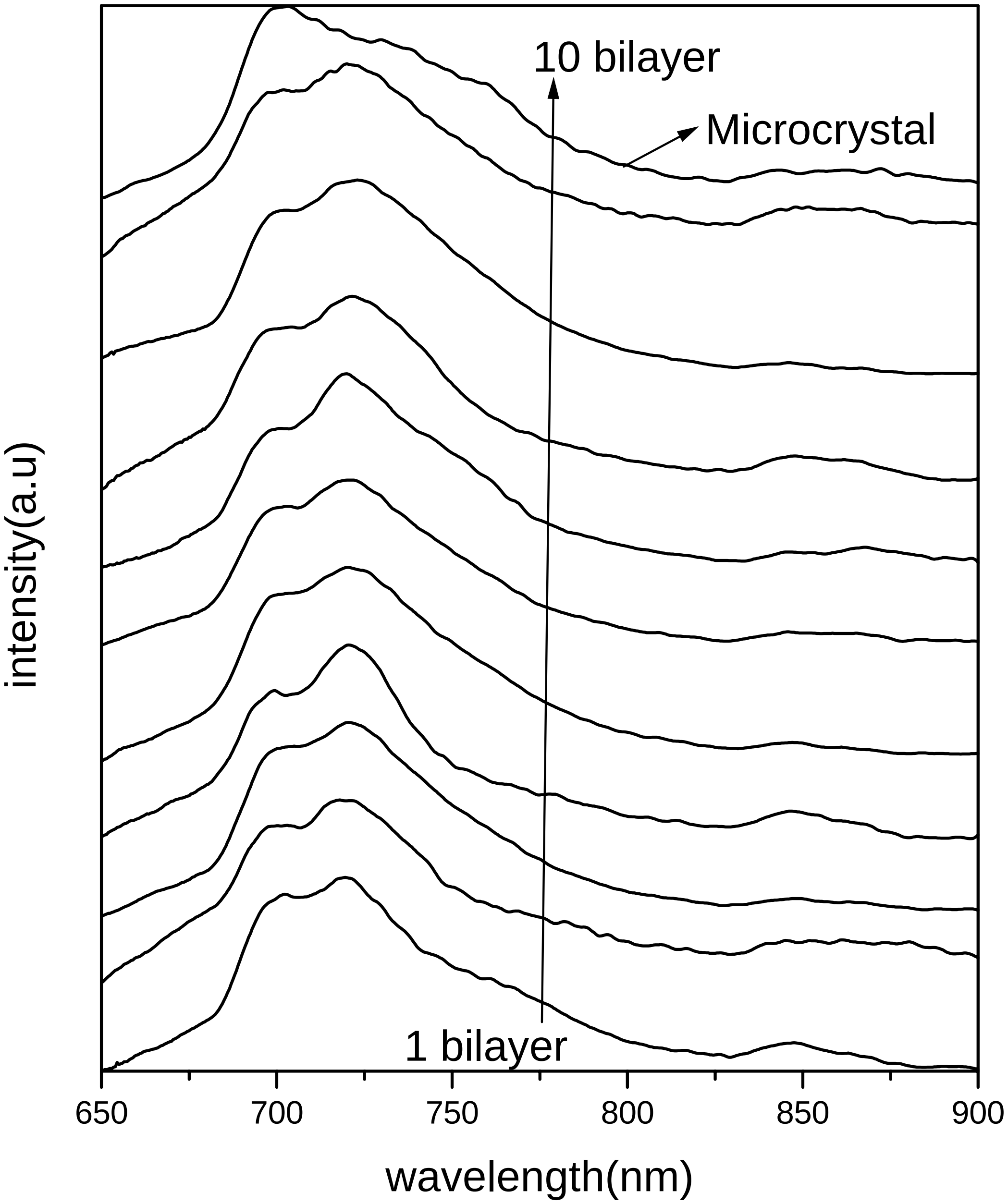

Figure 2b : Md. N. Islam et. al.

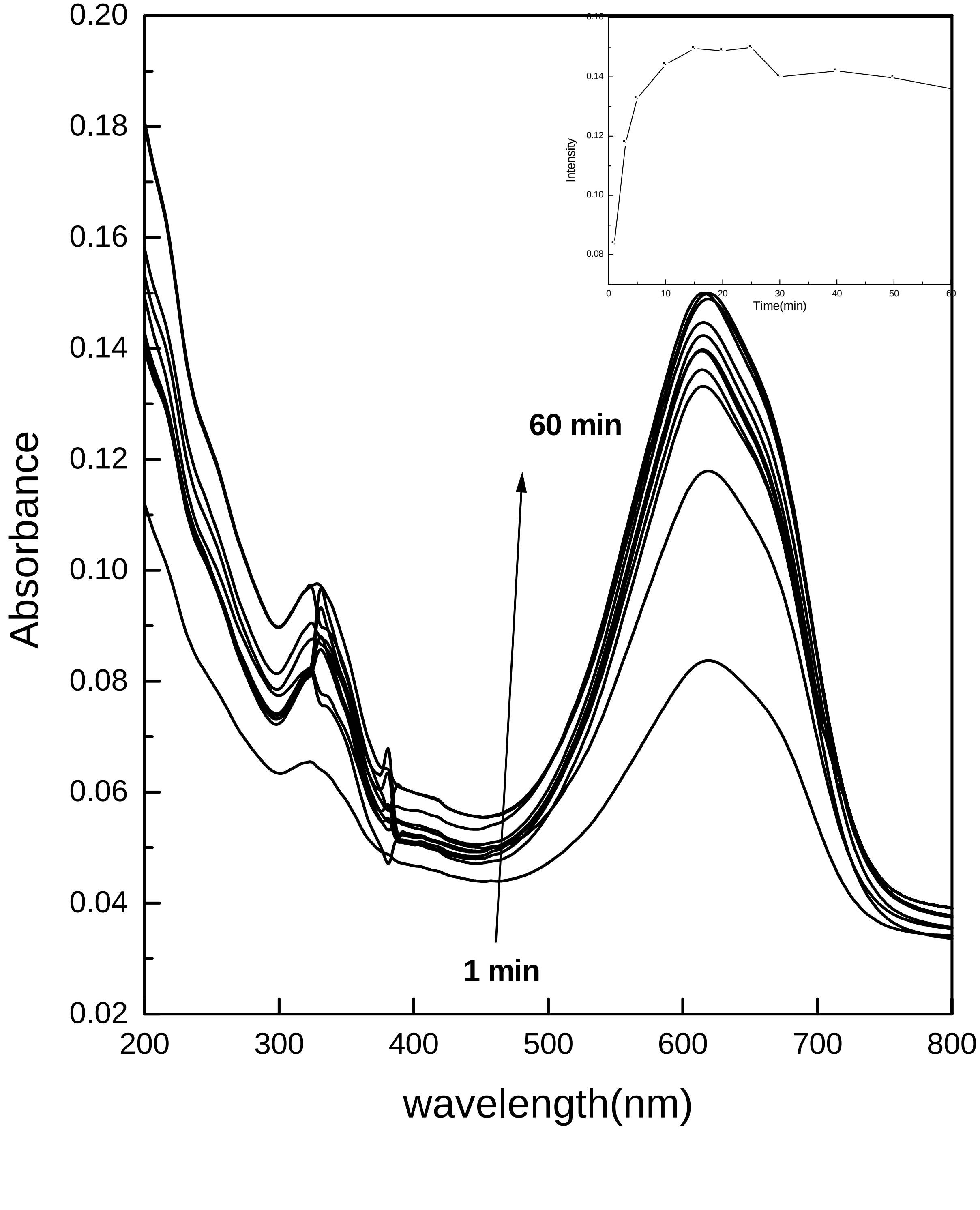

Figure 3a : Md. N. Islam et. al.

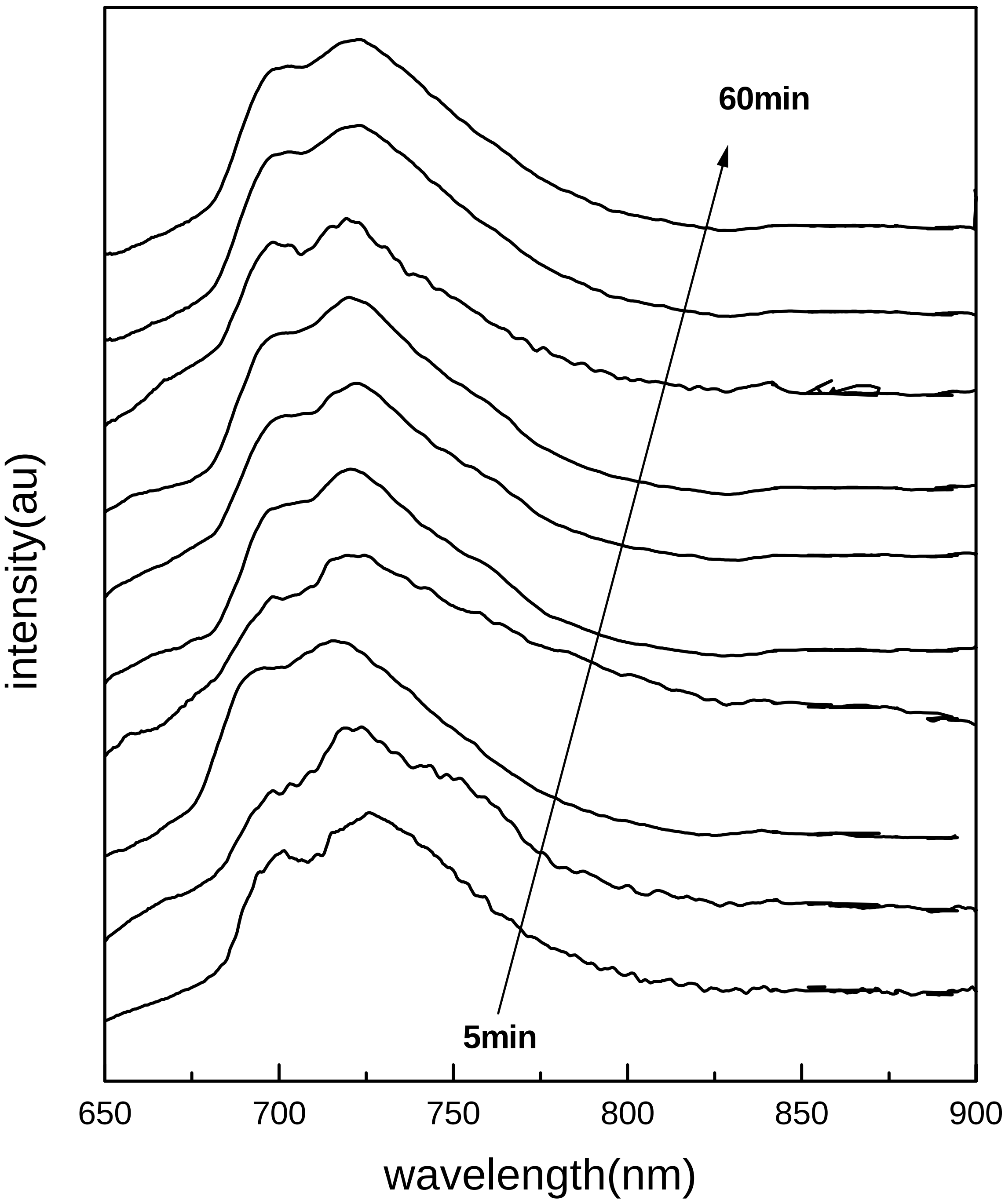

Figure 3b: Md. N. Islam et. al.

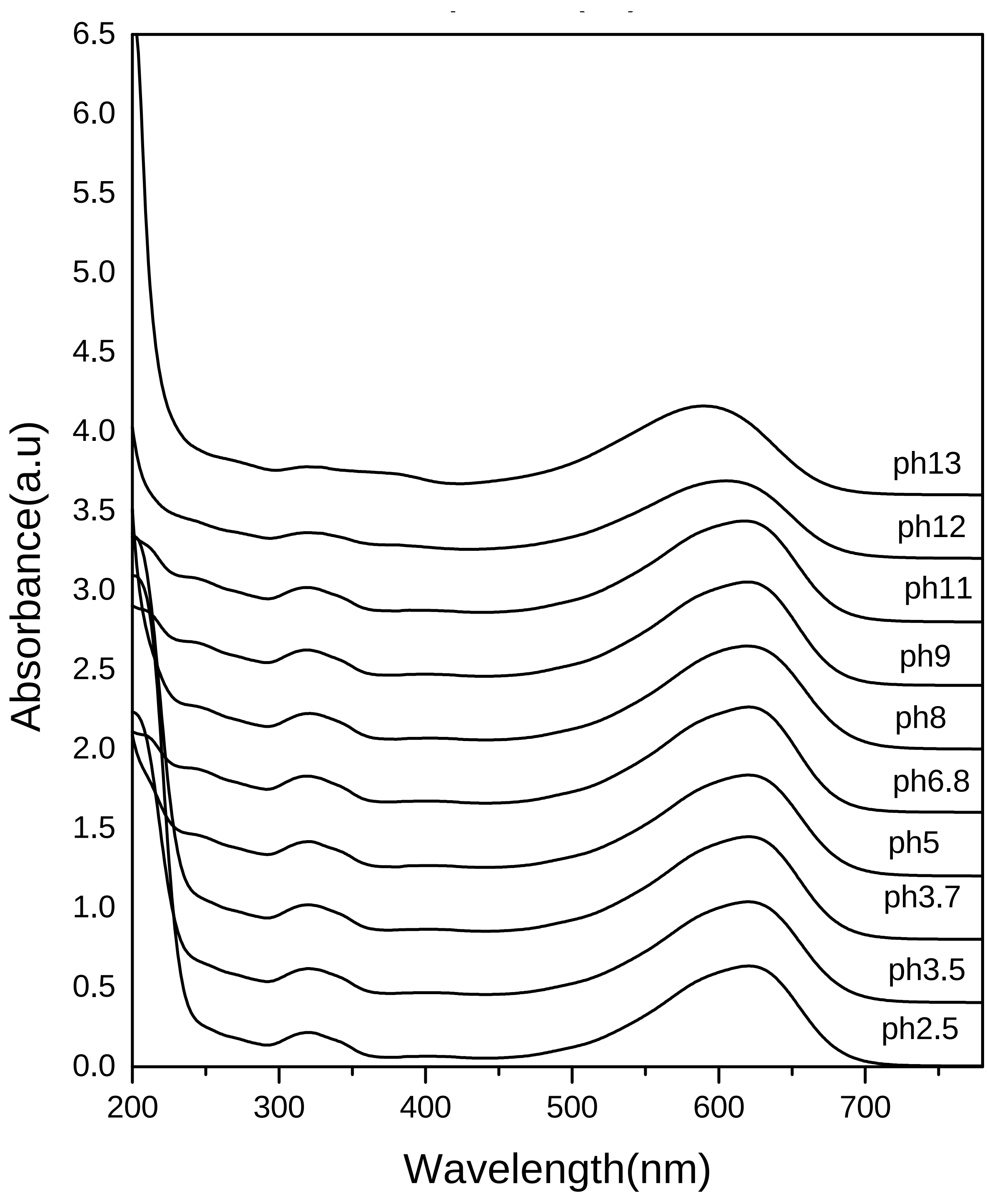

Figure 4a: Md. N. Islam et. al.

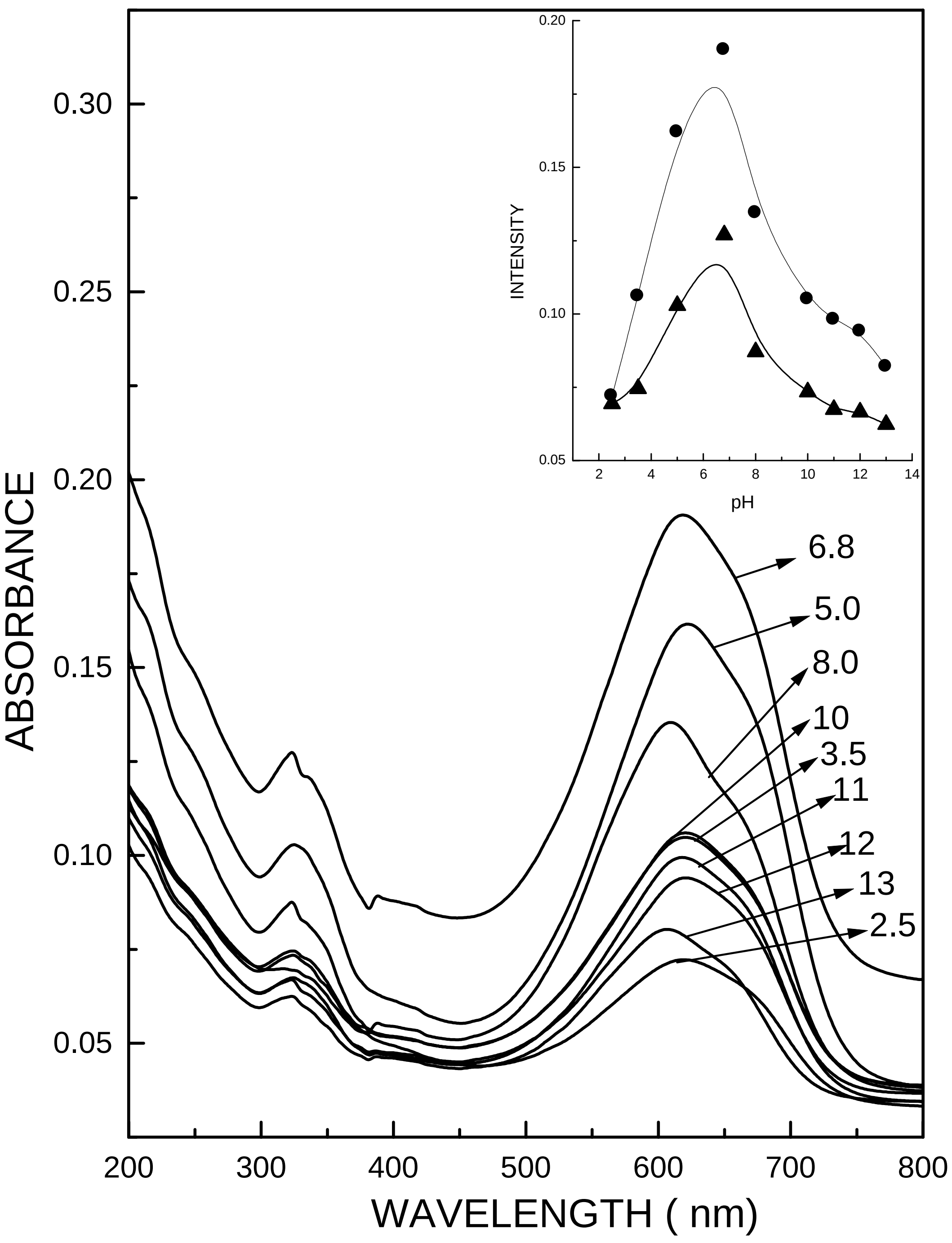

Figure 4b: Md. N. Islam et. al.

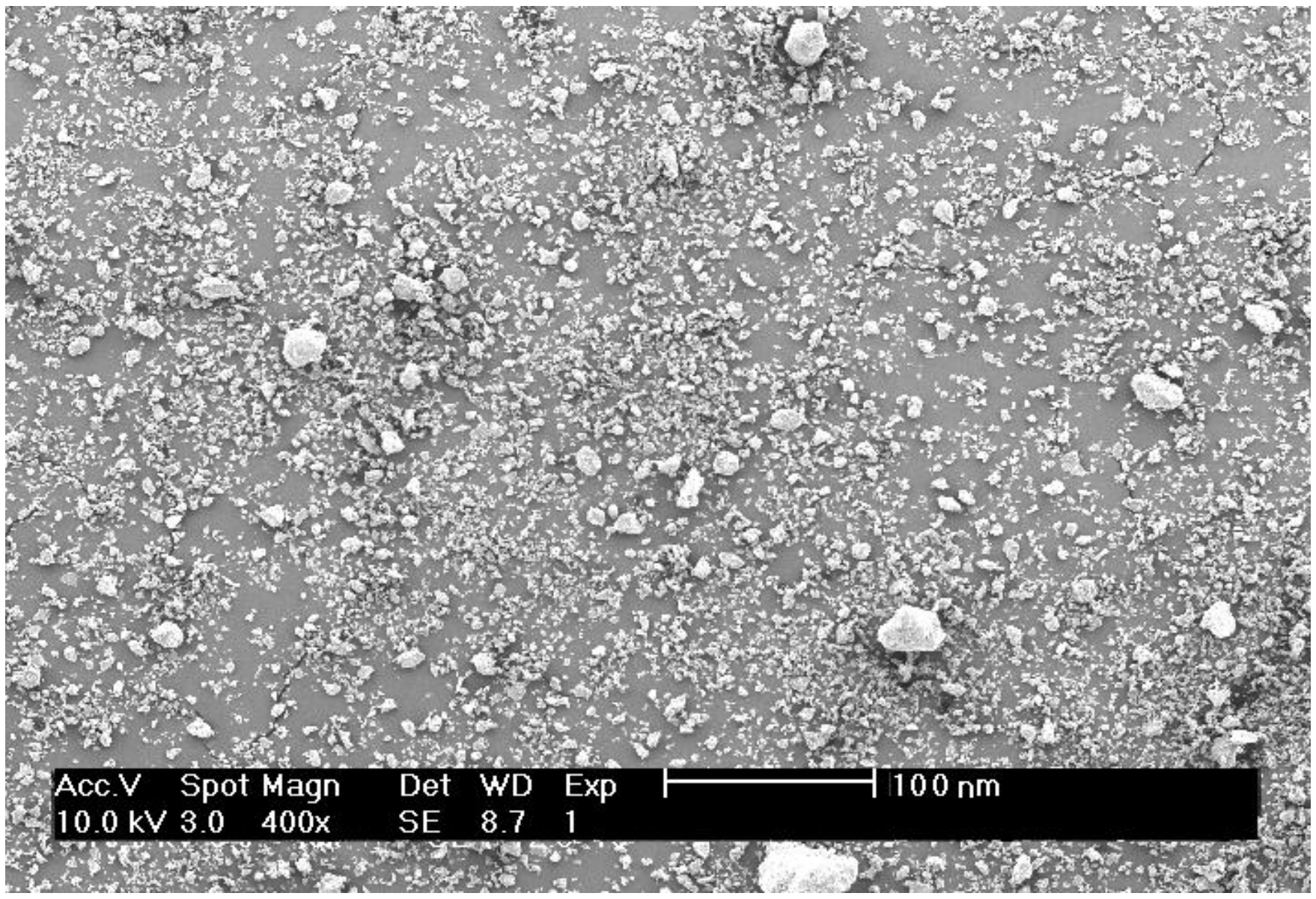

Figure 5: Md. N. Islam et. al.